# Structure Focused Neurodegeneration Convolutional Neural Network for Modelling and Classification of Alzheimer's Disease

Simisola Odimayo[1], Chollette C. Olisah[1,*], and Khadija Mohammed[2]

1 School of Engineering, University of the West of England, United Kingdom
2 Computer Science, Baze University, Abuja, Nigeria

*Correspondence to chollette.olisah@precximed.com, UWE, Coldharbour Ln, Stoke Gifford, Bristol BS16 1QY

**ABSTRACT**

Alzheimer's disease (AD), the predominant form of dementia, is a growing global challenge, emphasizing the urgent need for accurate and early diagnosis. Current clinical diagnoses rely on radiologist expert interpretation, which is prone to human error. Deep learning has thus far shown promise for early AD diagnosis. However, existing methods often overlook focal structural atrophy critical for enhanced understanding of the cerebral cortex neurodegeneration. This paper proposes a deep learning framework that includes a novel structure-focused neurodegeneration CNN architecture named SNeurodCNN and an image brightness enhancement preprocessor using gamma correction. The SNeurodCNN architecture takes as input the focal structural atrophy features resulting from segmentation of brain structures captured through magnetic resonance imaging (MRI). As a result, the architecture considers only necessary CNN components, which comprises of two downsampling convolutional blocks and two fully connected layers, for achieving the desired classification task, and utilises regularisation techniques to regularise learnable parameters. Leveraging mid-sagittal and para-sagittal brain image viewpoints from the Alzheimer's Disease Neuroimaging Initiative (ADNI) dataset, our framework demonstrated exceptional performance. The para-sagittal viewpoint achieved 97.8% accuracy, 97.0% specificity, and 98.5% sensitivity, while the mid-sagittal viewpoint offered deeper insights with 98.1% accuracy, 97.2% specificity, and 99.0% sensitivity. Model analysis revealed the ability of SNeurodCNN to capture the structural dynamics of mild cognitive impairment (MCI) and AD in the frontal lobe, occipital lobe, cerebellum, temporal, and parietal lobe, suggesting its potential as a brain structural change digi-biomarker for early AD diagnosis. This work can be reproduced using code we made available on GitHub.

**Introduction**

Alzheimer's disease (AD) is a specific type of dementia associated with severe neurological deficits that affect cognitive, visual, sensory, and motor functions in people living with the disease[1]. With AD, neurodegeneration, a progressive loss of structure or function of neurons, is inevitable, and there is currently no cure for reversing this process. However, clinical studies have shown that neurodegeneration progresses. With early diagnosis, treatment, and therapeutic interventions, the process can be slowed. At present, a definitive diagnosis of AD remains a complex task because tests for the presence of amyloid plaques and phosphorylated tau are the true determinants of AD and can mainly be performed posthumously[2]. Other clinical practices depend on a multitude of evaluations, including clinical assessments, medical history reviews, cognitive assessments, and neuroimaging, and many years of study are needed to reach a diagnostic decision[3]. Neuroimaging methods, such as positron emission tomography (PET) and MRI, provide information on the extent of structural changes in the brain relevant for pathological alterations characteristic of the brain during degeneration[4], particularly MRI. MRIs reveal several broad viewpoints, such as axial, coronal, and sagittal, with different levels of information for analysing brain neurodegeneration. Notably, the axial view revealed substantial atrophy of the cerebral cortex, leading to shrinkage of the outer layer of the cerebrum. This atrophy is accompanied by ventricle enlargement, reduced brain volume, and diminished gray matter[5]. In contrast, the coronal view highlights ventricle enlargement and emphasizes significant temporal lobe and cortical atrophy. This is a window into the widespread loss of neurons throughout the brain, accompanied by sulcus widening and gyrus thinning[5]. The sagittal plane provides the most visible information for AD diagnosis[7]. Brain neurodegeneration is evident in the sagittal plane, particularly in the frontal lobe, cerebellum, occipital lobe, thalamus, and corpus callosum, where learning and memory, mental function, motor function, and sensory function[7] can be significantly impacted.

Despite the diagnostic potential of MRI, its sole reliance on early AD diagnosis faces numerous limitations[9]. For example, AD may elude visual detection, especially when numerous samples of MCI and AD patients are analysed, thus necessitating the need for a comprehensive clinical evaluation methodology that is reliable. Additionally, the interpretability of MRI scans varies among radiologists and clinicians, which can introduce inconsistency in diagnosis. However, the emerging techniques in deep learning offer promise in diagnosing AD, particularly in timeliness, thereby paving the way for more effective AD management and intervention.

Over the past decade, deep learning algorithms, including both pretrained networks and tailored architectures, have been successfully adopted for AD modelling. Pretrained networks have long-standing relevance in AD diagnostic research. Bae tailored a residual network-50 (ResNet50)[10] for discriminating between MCI and AD patients and achieved an accuracy of 82.4%. The GoogLeNet, AlexNet, and ResNet-18 pretrained networks were exploited for classifying patients into cognitively normal, early mild cognitive impairment, mild cognitive impairment, and late mild cognitive impairment categories[11]. With accuracies of 96.39%, 94.08%, and 97.51% for GoogLeNet, AlexNet, and ResNet-18, respectively, ResNet-18 outperforms the other models in terms of performance. By integrating a 3D mobile inverted bottleneck convolution (MBConv) block in a 3D EfficientNet architecture[12], accuracy, sensitivity, specificity, and AUC values of 86.67%, 75.00%, 90.91%, 97.16%, and 83.33%, respectively, were achieved for the sMCI and pMCI sets. In another work, the DenseNet-169 and ResNet-50 CNN architectures were exploited for early AD diagnosis [13]. DenseNet-169 exhibited superior accuracy, surpassing ResNet-50, with scores ranging between 97.7% and 88.7%. The ResNet-18 pretrained network was useful for AD classification[14]. With the use of the Mish activation function (MAF) for enhancing the model's learning adaptability and a weighted cross-entropy loss function to ensure equitable consideration of the AD, MCI, and CN classes, the network achieved 88.3% accuracy on the preprocessed ADNI dataset.

Tailored deep learning algorithms are now paving the way for AD diagnosis. Basaiaa *et al.*[15] proposed a 3D CNN consisting of 2 convolutional blocks of $5 \times 5 \times 5$ filter sizes and 10 blocks of $3 \times 3 \times 3$ filter sizes. They utilized strides in place of max-pooling for downsampling. Their work achieved 74.8%, 75.1%, and 75.3% accuracy, sensitivity, and specificity, respectively, on the ADNI stable (s-MCI) and MCI conversion (c-MCI) sets. The model achieved 85.9%, 83.6%, and 88.3% accuracy, sensitivity, and specificity, respectively, on the AD and s-MCI sets. Another study proposed a CNN for AD diagnosis and stratification[16]. This research not only facilitated fast and accurate AD diagnosis but also offered classification for normal, MCI, and AD patients. Additionally, we addressed the challenging task of stratifying MCI into very mild dementia (VMD), mild dementia (MD), and moderate dementia (MoD) stages, akin to prodromal AD. The shallow network[16] achieved an overall testing accuracy of 99.68%, which was greater than that of pretrained networks such as DenseNet121, ResNet50, VGG 16, EfficientNetB7, and InceptionV3. Although the dataset used was the Open Access Series of Imaging Studies (OASIS) dataset, it is important to note that their work shows the importance of custom-trained networks in AD diagnosis. A fine-tuned CNN classifier[17] called AlzheimerNet was shown to be capable of classifying Alzheimer's disease into five stages. With data preprocessing and augmentation, their method achieved 98.67% accuracy using the RMSProp optimizer. Considering the different patient groups used for diagnosing AD, 18 independent binary classifications were proposed: healthy control (HC) vs. AD, HC vs. pMCI, HC vs. sMCI, pMCI vs. AD, sMCI vs. AD, and sMCI vs. pMCI according to the deep belief network (DBN). The modifications to the DBN[18] include dropout and zero-masking for overcoming overfitting, a preprocessing algorithm, a principal component analysis for dimensionality reduction, and a multitask feature selection approach. Using the ADNI dataset, accuracies ranging from 87.78% to 99.62% were observed. Hazarikar *et al.*[19] replaced the downsampling layer in the traditional LeNet architecture with a fusion of min-pooling and max-pooling layers to retain both minimum-value and maximum-valued signals. Their model achieved an accuracy, precision, recall, and F1-score of 98%, 96%, 97%, and 98%, respectively, on the ADNI dataset. In another work, a VGG-TSwinformer architecture[20] that combines a VGG-16 convolutional neural network and a transformer network was proposed and validated on the ADNI sMCI and pMCI cohorts. The accuracy, sensitivity, specificity, and AUC were 77.2%, 79.97%, 71.59%, and 0.8153, respectively. Similarly, another work[21] also found architecture useful for their methodology. They created a hybrid architecture by combining AlexNet with LeNet and varying the filter sizes from $1 \times 1$, $3 \times 3$, and $5 \times 5$. Scores as high as 96%, 93%, 93%, and 96% for accuracy, precision, recall, and F1 score, respectively, were reported. Another notable architecture is the multiplane convolutional neural network (Mp-CNN) architecture, which simultaneously processes three planes, axial, coronal, and sagittal, of 3D MRI [22]. The architecture of the Mp-CNN comprises 14 layers with rectified linear unit (ReLU) activation and softmax for multiclass classification, and it outperforms traditional 2D CNNs in multiclass classification associated with AD, MCI, and NC. The Swinformer has also been explored[23] as a transformer-based CNN architecture for AD classification. The Swinformer combines a CNN module for planar feature extraction and a transformer encoder module for 3D semantic connections. They argued that Swinformer can capture local features more accurately. The pipeline included data preprocessing and augmentation strategies such as random rotation and mirror reflection and recorded an accuracy of 88.3%.

While it is obvious that transfer learning with the use of state-of-the-art pretrained models is a promising technique for diagnosing AD, the tailored deep learning algorithm outperforms traditional methods in terms of performance and shows that it is better suited to preserving the underlying structure of the data for diagnosing AD. However, of these works, none have captured the structural dynamics of neurodegeneration in the brain in individuals with MCI or AD, which leaves room for additional work to be done. Further, the transfer-learning-based CNN architectures will fall short of their state-of-the-art performance expectations because the pre-trained weights are drawn from features unrelated to the focal structural atrophy of cerebral cortex neurodegenration. And since the other tailored-CNN architectures were not designed with considerations to the focal structural atrophy of the cerebral cortex neurodegeneration, their performance will be comparable to the pre-trained models. Therefore, this paper seeks to bridge this research gap by proposing SNeurodCNN. Unlike existing deep learning architectures, the SNeurodCNN architecture takes as input the focal structural atrophy features resulting from the segmentation of the structures of the cerebral cortex of the brain captured through magnetic resonance imaging (MRI). This is the first time the focal structural atrophy feature is investigated for its relevance in AD diagnosis. Following are the contributions of this paper to AD classification research.

- We propose a framework that integrates a preprocessor comprising of the novel SNeurodCNN architecture for modelling the structural neurodegeneration of the brain's cerebral cortex and for the task of discriminating between MCI and AD. The architecture considers only necessary CNN components, which comprise two downsampling convolutional blocks and two fully connected layers, and utilises regularisation techniques to regularise learnable parameters to achieve state-of-the-art performance in AD classification. We show that SNeurodCNN is sufficient and better at learning the structural changes of the cerebral cortex resulting from brain neurodegeneration better than with pre-trained models. The SNeurodCNN performance in the deep learning framework is further enhanced through the expansion of pixel intensity brightness using the Gamma correction technique.
- By leveraging the mid-sagittal and para-sagittal brain image viewpoints of the Alzheimer's Disease Neuroimaging Initiative (ADNI) dataset, we investigate whether the varying viewpoints of the two planes of the sagittal axis, midsagittal and parasagittal, provide differing SNeurodCNN insights into structural neurodegeneration.
- We investigate SNeurodCNN sensitives to brain neurodegeneration for perceiving important digital biomarkers (digi-biomarkers) in order to identify facets of the brain where focal structural neurodegeneration of the cerebral cortex is prevalent.

**Results**

This section presents the results and findings of this paper. We begin by outlining the findings of SNeurodCNN and then progress to its analysis using Grad-CAM to identify features strongly indicative of the model's sensitivities to brain neurodegeneration.

Evaluation of the Structure-focused ADNI Dataset

Experiments on the structure-focused ADNI data version of the brain are relevant for understanding the structural changes that are contributors to brain neurodegeneration. The performance of the SNeurodCNN was evaluated on the midsagittal and parasagittal planes of the sagittal axis of the brain. The parasagittal view of the sagittal plane is the version taken from the off-centre plane, while the midsagittal view is the version taken at exactly the centre of the plane. The use of both planes is to investigate whether the model is impacted differently by neurodegeneration, as it pertains to the parasagittal region, which exposes regions of motor control, sensation, and perception, and the midsagittal region, which exposes regions that are responsible for visuospatial integration, memory, and self-awareness. The results of this experiment are presented in Table 1 and Fig. 1 (a) and show that SNeurodCNN performed equally well with the varying viewpoints of the two planes of the sagittal axis, midsagittal and parasagittal, though the mid-sagittal plane provided better insight into brain neurodegeneration than the para-sagittal plane. Fig. 1 (b) illustrates the significance of enhancing pixel brightness. A performance increase of 2.4 % is achieved with SNeurodCNN with Gamma correction compared to without correction.

Table 1: Performance evaluation of SNeurodCNN classification model in the midsagittal and parasagittal planes

| Metric | Midsagittal Plane | Parasagittal Plane |
| --- | --- | --- |
| Accuracy | 98.1% | 97.8% |
| Precision | 97.2% | 97.2% |
| Sensitivity | 99.0% | 98.5% |
| Specificity | 97.2% | 97.0% |
| F1-score | 98.1% | 97.8% |
| AUC | 98.1% | 97.8% |

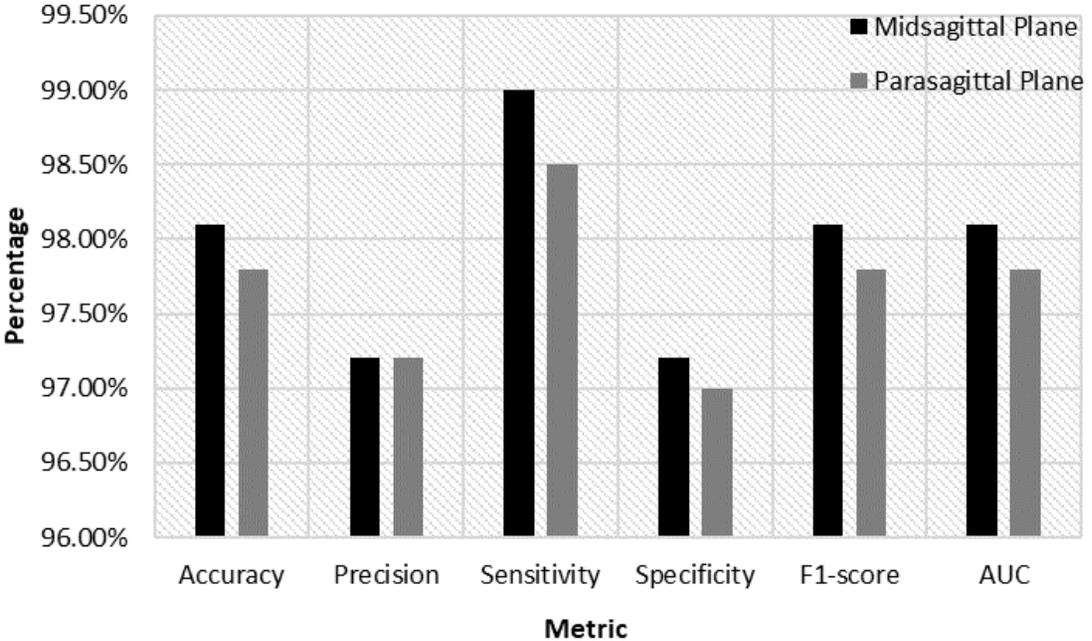

(a)

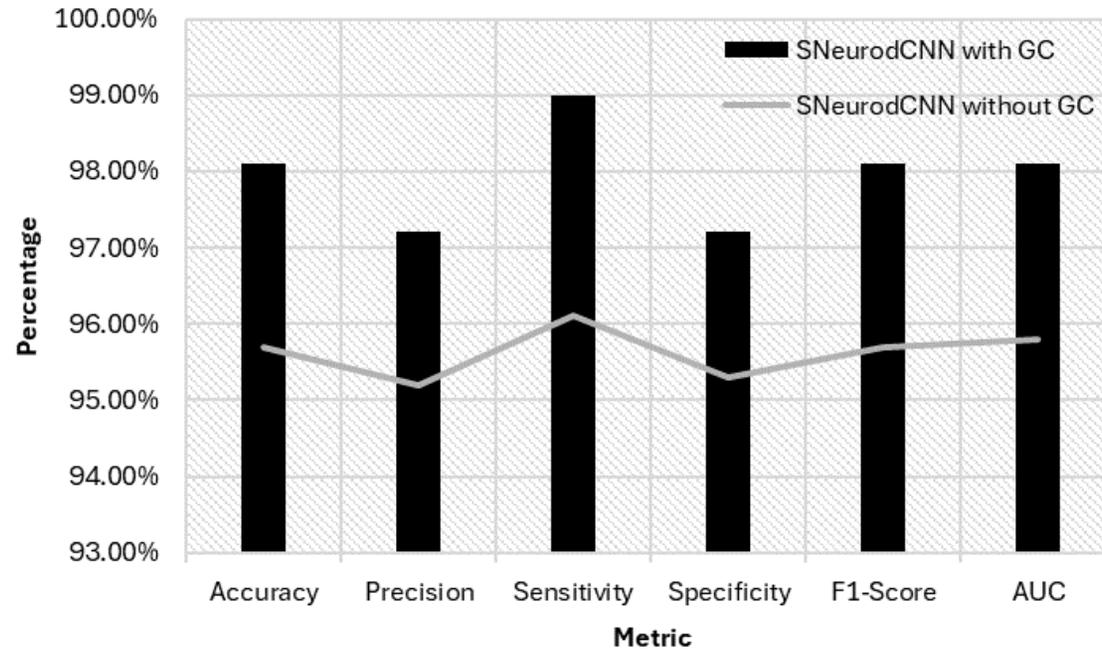

(b)

Fig. 1: Classification performance of SNeurodCNN model. The results combine performances at the midsagittal and parasagittal planes. Performance improvement with Gamma correction using the mid-sagittal plane.

For a detailed view of performance during training, we present the training and validation loss and accuracy generated during SNeurodCNN model development to better judge the model's generalization capability on the unseen set. The results can be observed in Fig. 2. the training and validation loss and accuracy performance of SNeurodCNN with training sets from the midsagittal and parasagittal viewpoints. The loss curves generated in Fig. 2 (a) and (b) demonstrate an exceptional trend in the training progress for both the parasagittal and midsagittal regions of the brain. As the number of epochs increases, there is a smooth and consistent descent in the loss curves, indicating effective model training. The convergence of training and validation losses to stable values, with a plateau at higher loss levels, signifies successful learning without succumbing to overfitting. Additionally, the parallel curves of both validation and training further attest that SNeurodCNN is learning from the training data but also demonstrates a robust ability to generalize to previously unseen data and simultaneously identify intricate patterns in the data while avoiding the pitfalls of overfitting.

Further, akin to the loss curves generated by the model, the accuracy curves depicted in Fig. 2 not only demonstrate the model's adeptness in making predictions for both datasets but also showcase its adaptability to the training data, indicating a continuous enhancement in predictive accuracy. As the epochs increase, the upward trajectory of the axis curve for both experiments corresponds to the decrease in loss. Although peak accuracy is reached relatively early on for both the midsagittal and parasagittal datasets, this does not compromise the SNeurodCNN performance as the close alignment between training and validation indicates good generalization without overfitting.

To appear in Scientific Report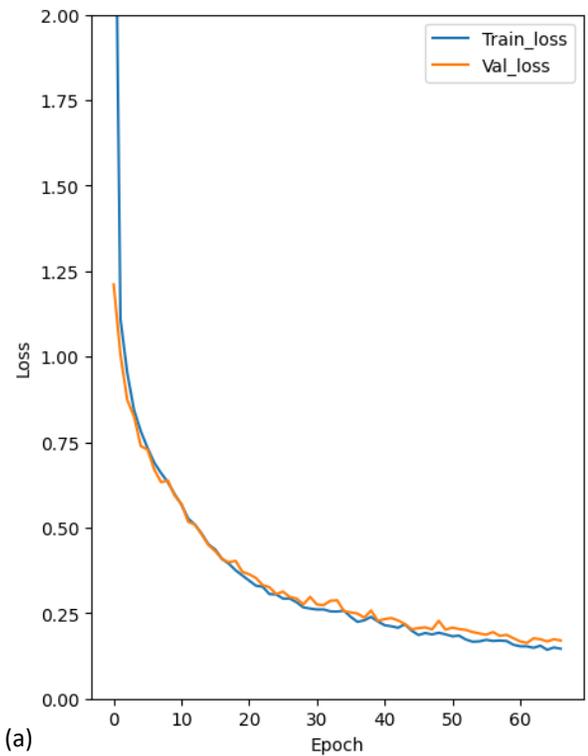
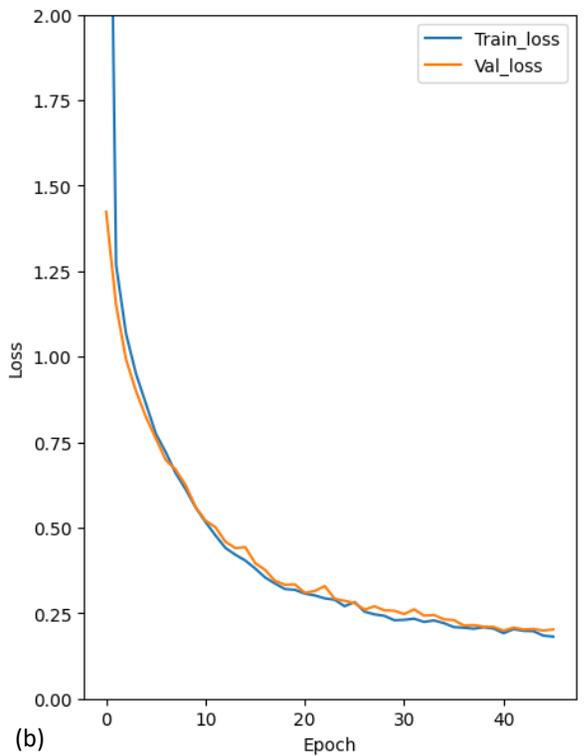

(a) (b)

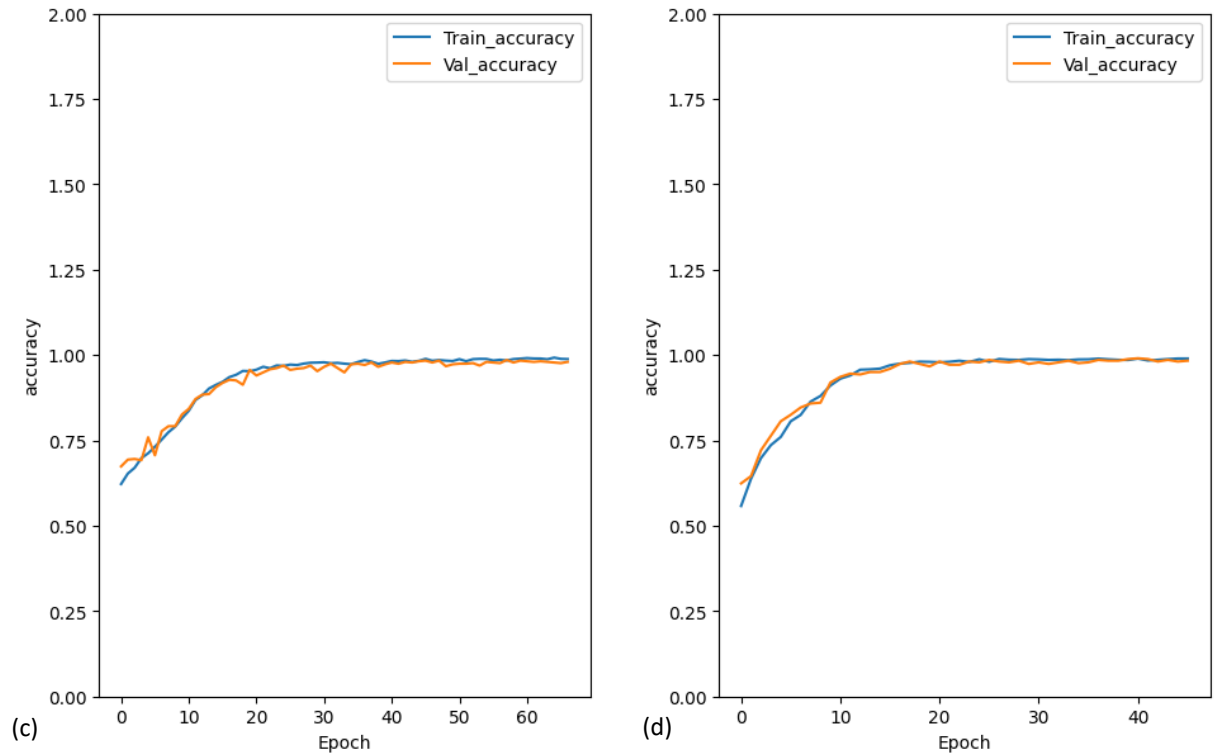

Fig 2: Plot of the training and validation loss and accuracy curves generated during the learning process. Graphs (a) and (c) depict the graphs achieved by using parasagittal region-focused images while (b) and (d) depict the graphs achieved by the midsagittal region.

As shown in Table 1 and Fig. 1, the accuracy, precision, recall, specificity, F1 score, and AUC of the SNeurodCNN are outstanding. This is an indication that the model is capable of modelling the structural neurogenerative impacts of MCI and AD occurring at both parts of the sagittal plane—the midsagittal and parasagittal planes. While the accuracies recorded are high, which shows the model's ability to distinguish between instances of AD and instances of MCI, the midsagittal accuracy was increased by 0.3%. Given that the structure-focused ADNI is used, the performance conforms with the medical statement "midsagittal cerebral morphology provides a homologous geometrical reference for brain shape and cortical vs. subcortical spatial relationships" [24]. In medical diagnosis, a high recall is preferred—the higher the value is, the greater the confidence we have in the model's ability to minimize false positives, that is, diagnosing AD when in fact it is MCI and vice versa. Furthermore, the high F1-scores for both models reflect well-balanced performance in terms of precision and recall, ensuring a lower rate of both false-negative and false-positive predictions.

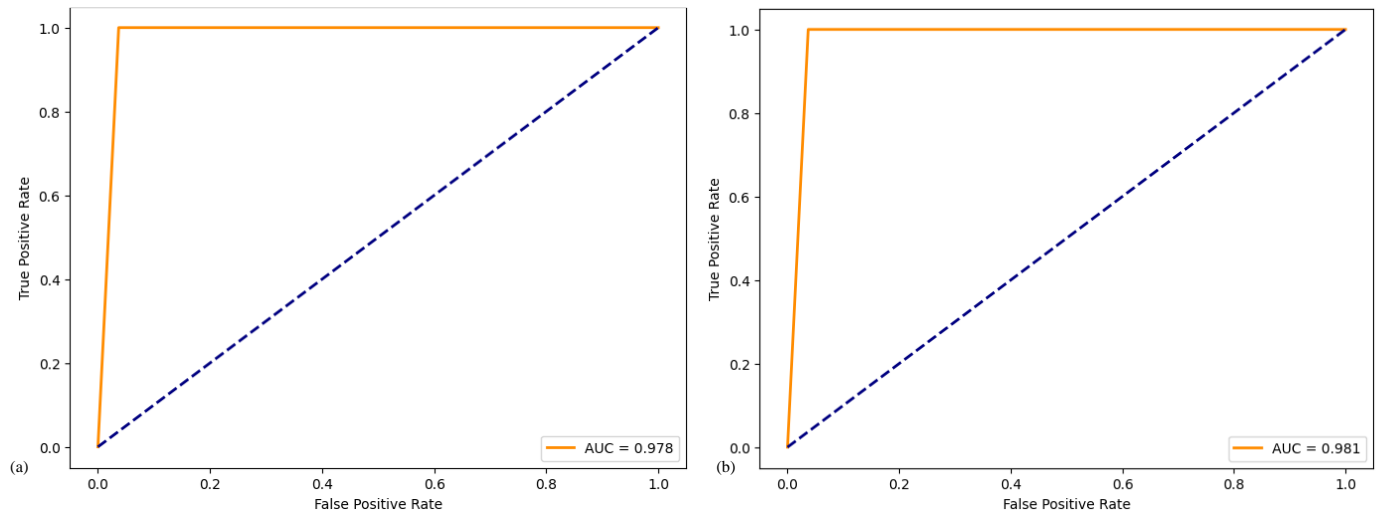

Fig. 3: Plot of the Receiver Operating Characteristic for Parasagittal dataset (a) and Midsagittal dataset (b). The ROC curves illustrate the discrimination of the classification model on the respective dataset while showcasing the model's excellent trade-off between sensitivity and specificity with a steep ascent and high AUC scores. These high AUC values of 0.978 (97.8%) and 0.981 (98.1%) further demonstrate the model's superior predictive accuracy in identifying AD and differentiating between MCI patients.

As an additional measure of our model's performance, we utilized the AUC metric, which is a metric that provides an unbiased view of a model's performance amidst class imbalances aside from the F1 score since the data presented are 180:105 in proportion for AD patients and MCI patients. Scores of 98.1% and 97.8% were observed for the midsagittal and parasagittal sides, respectively, which underscores the models' discriminatory capability in diagnosing AD.

SNeurodCNN Sensitivity to Brain Neurodegeneration

To understand the sensitivity of SNeurodCNN to brain neurodegeneration, the same slices from the MCI and AD categories for the midsagittal and parasagittal planes were sampled. Then, with Grad-CAM, the activation maps of the last convolutional layer of SNeurodCNN are visualized to better understand the network's sensitivity to brain degeneration. The Grad-CAM output was analysed using a heatmap; the brighter the yellow, the more significant the region is, and the more prominent the region is where brain neurodegeneration is. The closer to purple it is, the less significant the difference is. These regions can be visualised in Fig. 4, with more examples for the midsagittal in Fig. 5 and parasagittal in Fig. 6.

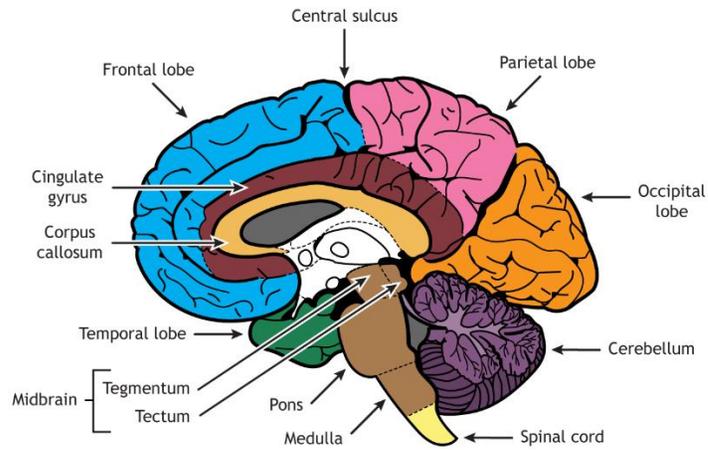 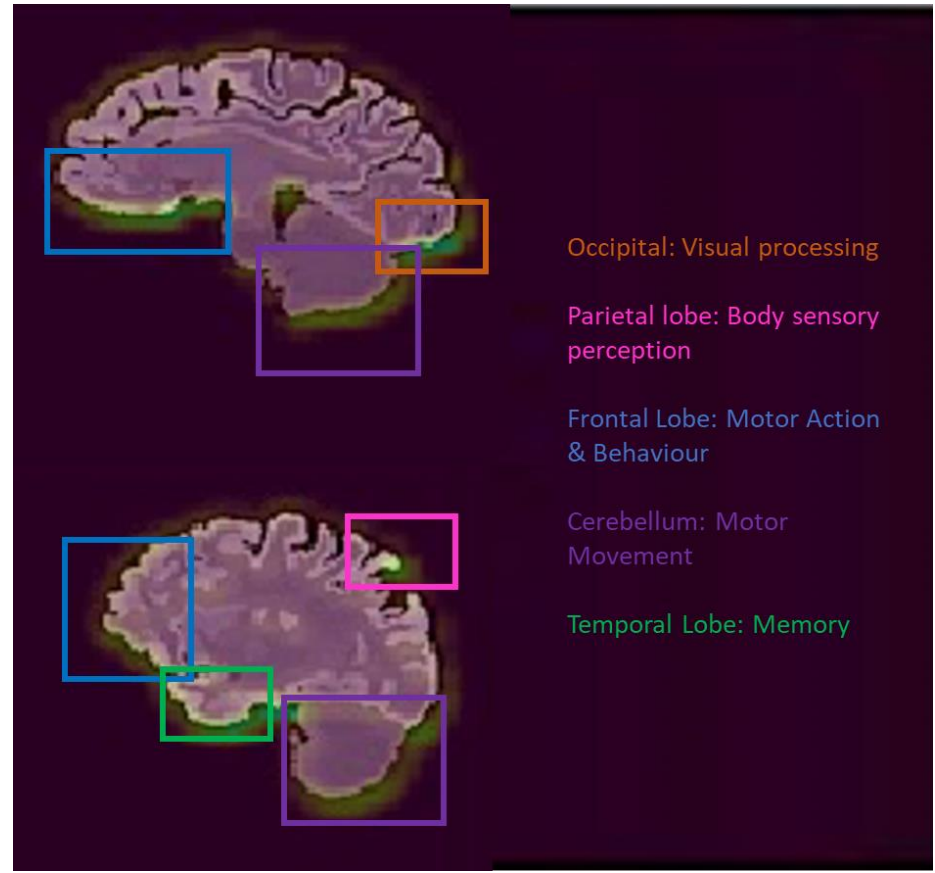

(a)           (b)

Fig. 4: Regions of brain neurodegeneration. (a) Midsagittal brain internal regions (Source: 'Internal Brain Regions' by Casey Henley, licensed under CC BY-NC-SA 4.0 International License, (b) Highlights of SNeurodCNN neurodegeneration sensitivity in the midsagittal (up) and parasagittal (down).

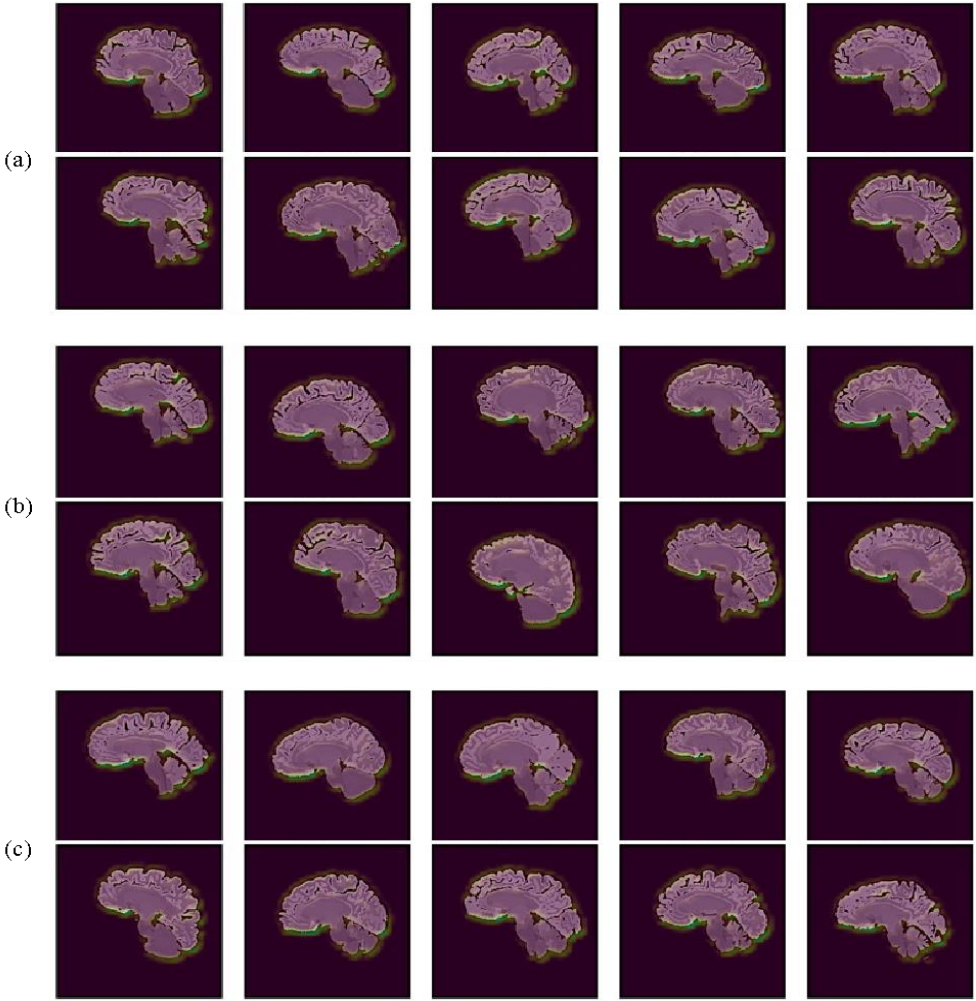

(a)

(b)

(c)

Fig. 5: Visualization of regions of interest on the midsagittal plane identified by the CNN model in AD (a), pMCI (b) and sMCI patients (c) subjects.

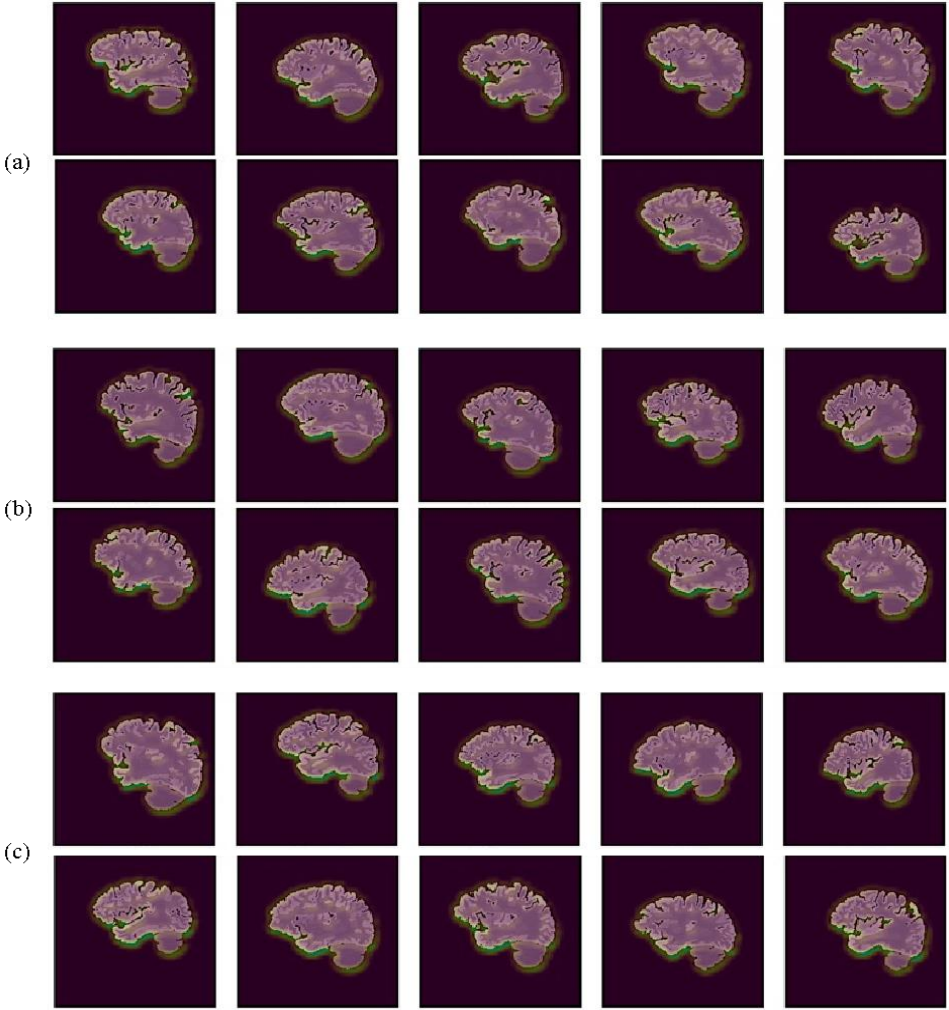

Fig. 6: Visualization of regions of interest on the parasagittal plane identified by the CNN model in AD patients (a), pMCI patients (b), and sMCI patients (c) subjects.

As can be observed from the MCI (sMCI and pMCI) and AD slices in Fig. 5, the frontal lobe, occipital lobe, and cerebellum are the regions highlighted as being highly significant, with the frontal lobe being the most prominent while the cerebellum the least prominent. Therefore, these regions reveal SNeurodCNN sensitivity to brain neurodegeneration in relation to AD. The findings made with the midsagittal plane are consistent with those made with the parasagittal plane and can be observed in Fig. 6, although the parasagittal plane shows additional prominent regions about the locations of the temporal and parietal lobe. Since we used the structure-focused ADNI, it is only expected that the highlights relate to changes in structure caused by shrinkage of the cerebral cortex, known as cerebral atrophy. An obvious interpretation of these observations is that the SNeurodCNN can sense a significant difference in the structural characteristics of the frontal lobe, occipital lobe, and cerebellum and possibly the parietal regions in individuals with MCI (sMCI and pMCI) and AD.

The succeeding discussions provide insights into the clinical relevance and functions of the frontal lobe, occipital lobe, cerebellum, temporal, and parietal lobe to better understand the roles they play in brain neurodegeneration in relation to AD. The frontal lobe, occipital lobe, and cerebellum are observed from the mid-sagittal viewpoint, while the temporal and parietal lobe are observed from the parasagittal viewpoint. These regions were well perceived by the SNeurodCNN as the digi-biomarkers for AD diagnosis and the intriguing thing is that the discovery corroborates with clinical studies.

- The frontal lobe is mainly responsible for motor action and the temporal integration of behavior [25]. The frontal lobes of patients with MCI show cortical atrophy[26] and severe cortical volume deficit in AD patients.
- The occipital lobe [27] is the region associated with visual processing for depth perception, color determination, object and face recognition, and is also responsible for memory formation. Early progressive MCI [28] manifests as structural change in the occipital lobe.

- The cerebellum is responsible for regulating motor movement and controlling balance[29]. Though it is not generally acknowledged, some degree of functional connectivity disruption can occur in the cerebellum of MCI and AD patients[30]. The sensitivity analysis with the proposed SNeurodCNN confirms this discovery because MCI and AD are observed to show a subtle structural impact on the cerebellum as shown in Fig. 4.
- The parietal region is responsible for body sensory perception. This region has not been confirmed in any of the DL-based studies, but a clinical study[31] shows that the volume atrophy of grey matter of the brain can indicate a difference between MRI of MCI patients from AD patients.
- The temporal lobe is responsible for memory loss. Memory loss, to a considerable extent, affects MCI and AD patients, with the former being less pronounced than the latter [32,33].

Comparison with the state-of-the-art methods

We first compare results generated with this paper's focal structural atrophy ADNI data using pre-trained ResNet50 and DenseNet169 models which were shown in the reviewed literature to achieve better performance compared to other pre-trained models. As can be observed in Table 2 and Fig. 7, the SNeurodCNN achieves a 32.6 % and 16.5 increase in accuracy compared to ResNet50 and DenseNet169 pre-trained models, respectively. This performance shows that SNeurodCNN better models the structural brain neurodegeneration because it is designed to consider only the convolutional blocks needed to encode the focal structural atrophy. Invariably, the structure of SNeurodCNN helped it to overcomes overfitting challenges compared to the deeper models. However, we are not affirming that the performance of the pre-trained networks is a lack of robustness but might be attributed to how less specialized they are at solving structural neurodegeneration problem. Further, we compare the reported accuracies of models from existing literature, the tailored and pre-trained architectures, where the focus is on AD and MCI classification with the ADNI dataset. Though these are in-direct comparisons as they were lifted *as it is* from the literature, the dataset and task of discriminating between MCI and AD make them relevant. The tailored architecture models, Basaiaa *et al.*[15] and the DBN model[18] offer general approaches for AD and MCI classification and achieved accuracies of 88.3 % and 99.62 %, respectively. The Mp-CNN model[22] utilizes multi-plane processing for MCI and AD classification and achieves 93 % accuracy. On the other hand, the accuracies of the pre-trained models as reported in Table 3 confirm they are less specialized to solving brain neurodegeneration problems. The combination of DenseNet-169 and ResNet-50 CNN models[16] achieved 83.82% accuracy on MCI and AD classification. The hybrid LeNet-AlexNet Model[21] performed better than the hybrid of DenseNet and ResNet, it was short of the best-performing tailored architecture model. While each model has distinctive strengths for diagnosing AD and MCI, SNeurodCNN stands out for its specific focus on the regions where structural neurodegeneration occurs, making it a significant model in the field.

Table 2: Comparison of the SNeurodCNN model with pre- trained models

| Metric | Model | | |
|---|---|---|---|
| | SNeurodCNN | ResNet50 | DenseNet169 |
| Accuracy | 98.1 | 65.5 | 81.6 |
| Precision | 97.2 | 62.1 | 82.4 |
| Sensitivity | 99.0 | 75.8 | 79.2 |
| Specificity | 97.2 | 55.6 | 83.7 |
| F1-Score | 98.1 | 68.3 | 80.8 |
| AUC | 98.1 | 65.7 | 81.5 |

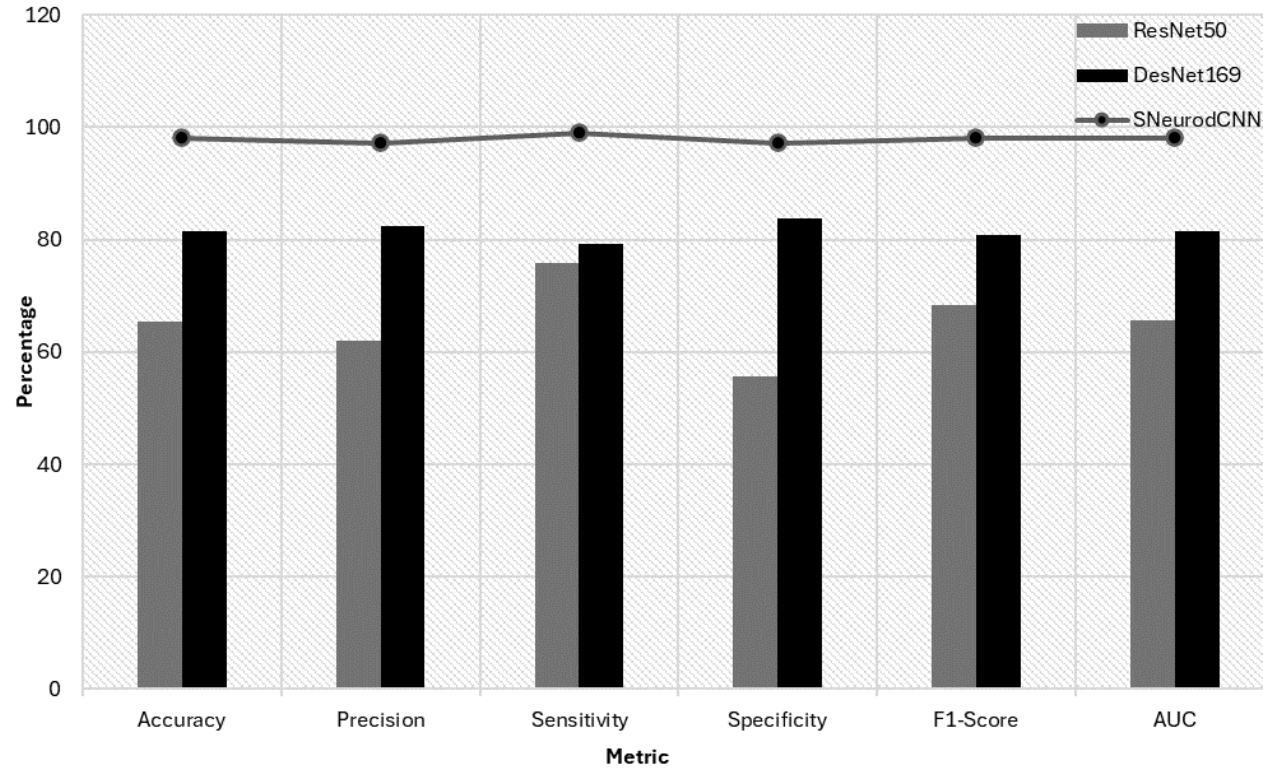

Fig. 7. Visualising the performance of SNeurodCNN in comparison to the pre-trained models.

Table 3: Comparison of the SNeurodCNN model with AD diagnosis state-of-the-art models in the literature

| S/N | Model | Year | Performance Metric | Architecture | Diagnostic Approach | Digi-Biomarker |
|---|---|---|---|---|---|---|
| 1 | Basaia et al.'s 3D CNN Model[15] | 2019 | 74.8% - 88.3% accuracy | 2 blocks of 5x5x5 and 10 blocks of 3x3x3 filters, use | AD and MCI diagnosis | - |
| 2 | Deep Belief Network (DBN) Model[18] | 2023 | 87.78% - 99.62% accuracy | strides for downsampling Dropout and zero-masking, multitask feature selection | Classification tasks for AD and MCI | - |
| 3 | DenseNet-169 and ResNet-50 CNN Models[16] | 2023 | DenseNet-169: 83.82% (testing); ResNet-50: Lower performance | Deep architectures with advanced feature learning | Early AD diagnosis | - |
| 4 | LeNet and AlexNet[21] | 2023 | 96% accuracy | Combines both models and applies varying filter sizes from $1 \times 1$, $3 \times 3$, and $5 \times 5$. | AD and MCI classification | - |
| 5 | Multiplane Convolutional Neural Network (Mp-CNN)[22] | 2022 | 93% accuracy | 14-layer architecture processing axial, coronal, and sagittal planes of 3D MRI | Advanced multiplane image processing | - |
| 6 | The proposed | 2024 | SNeurodCNN 98.1% accuracy 97.8% accuracy | Sagittal planes; CNN architecture: 2D Conv, 3x3 filter of varying depth, Maxpool, Dense layers, Dropout | Classification tasks for AD and MCI | Captures structural brain neurodegeneration about the frontal lobe, occipital lobe, cerebellum, temporal, and parietal lobe. |

**Discussion**

We delve into the discussion to elaborate on the implications of the results of this study. Neurologists traditionally use neuroimaging methods, such as MRI, to assess MCI and AD neurodegeneration, but this paper presents a dimension that automates diagnosis toward classifying the disease stages but can help in identifying the regions susceptible to neurodegeneration. This finding substantiates the relevance of the proposed SNeurodCNN model in the early diagnosis of AD. The evaluation of neurodegeneration using the midsagittal and parasagittal planes of the brain results in remarkable performance. These findings surpass those of previous studies[7,10]. SNeurodCNN has shown promise in capturing neurodegenerative features from MR images. As observed with Grad-CAM, structural changes were prominent in the frontal lobe, occipital lobe, and cerebellum on the midsagittal MR images. Frontal lobe dysfunction has been implicated in various neurodegenerative conditions, including AD and MCI[34]. The parasagittal view showed that an additional structural change was possible in the parietal lobe, which was not obvious in the midsagittal slices. The SNeurodCNN model region is most sensitive to the frontal lobe, occipital lobe, cerebellum, and parietal lobe, which is consistent with the findings of clinical studies [25-31] and, in some cases, consistent with regions identified in one of the DL-based studies[7]; however, our study further confirms the significance of cerebral atrophy on brain structure. Therefore, these findings show that our proposed SNeurodCNN can significantly contribute to AD diagnosis and prognosis. This necessitates the integration of deep learning into healthcare practices to provide early cues about AD that might be present in MCI patients, minimizing the occurrence of false positives and false negatives that can mitigate misdiagnosis.

Further, the structural changes that our proposed SNeurodCNN highlights set a record as a potential brain structural change digi-biomarker for the early diagnosis of AD. The benefits of digi-biomarkers are numerous. 1) These findings can help clinicians develop targeted and personalized treatment plans and interventions through personalized analysis of disease states. These findings could lead to valuable targets for therapeutic interventions and monitoring and the assessment of novel treatments accordingly.

**Materials and Method**

This section elucidates the dataset employed in this study and outlines the methodologies encompassing the data preprocessing pipeline and the design of the deep convolutional neural network architecture as illustrated in Fig. 8.

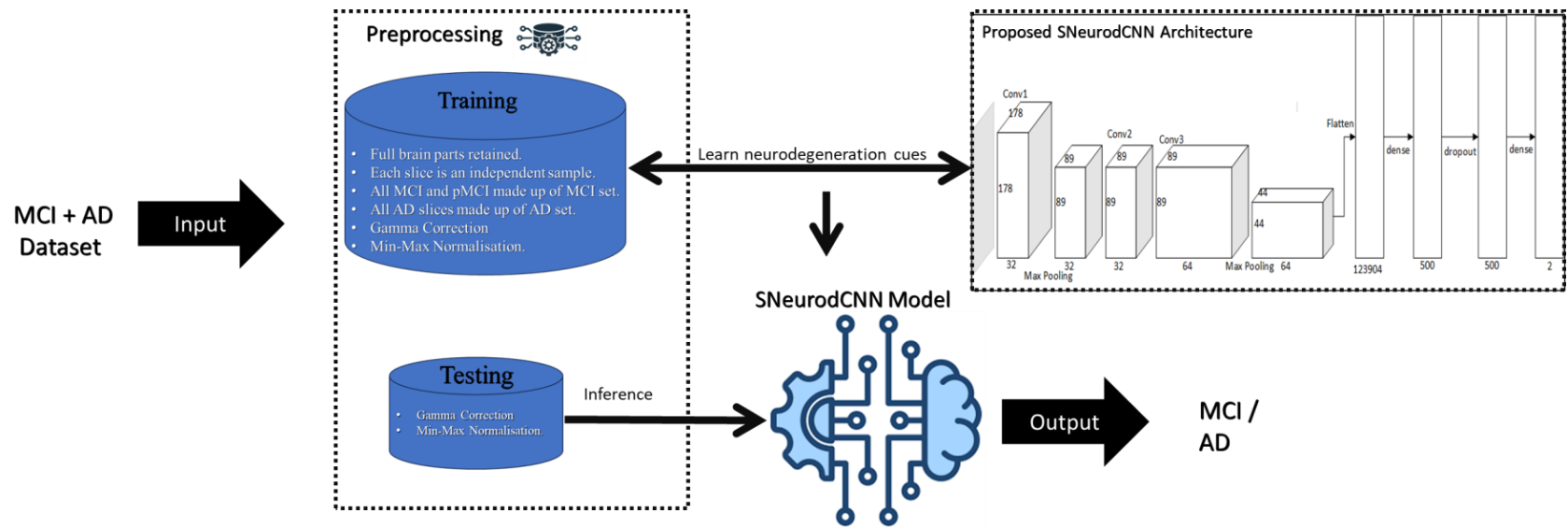

Fig. 8. Overview of the proposed deep learning framework for AD diagnosis.

Data

The ADNI AD benchmark dataset commonly used for AD analysis is relevant to this study. This dataset was designed from a study in which individuals aged 55 to 90 years were enrolled from 57 sites across the U.S. and Canada after providing informed consent. In all, the ADNI comprises 368 participants, categorized into 180 AD patients (82 females, 98 males; age ± SD = 75.28 ± 7.57 years) and 105 stable sMCI patients (41 females, 64 males; age ± SD = 74.69 ± 7.41 years). The average mini-mental state examination (MMSE) scores were 23 for AD patients, 28 for sMCI patients, and 27 for pMCI patients. All subjects had a T1-weighted baseline from the ADNI1/Go/2 cohort. The sMCI group consisted of individuals who were diagnosed with MCI at baseline and remained so for at least two years. A summary of the demographics of the ADNI participants is provided in Table 4 and the mental examination score of the participants in Fig. 9.

We particularly explored skull-free patient version. This version utilized the ADNI dataset preprocessed to correct imaging distortion through grad warping via *gradient inhomogeneity correction*, intensity correction, and scaling to address gradient drift. The results are further processed using multi-atlas label propagation with expectation maximization (MALPEM) to segment, on local scales, the cross-sectional structural volume changes in the brain, which constitute the structure-focused ADNI dataset. It should be noted that each slice of the structure-focused ADNI dataset is characterized by the region of the gray matter that forms the anatomical structures of the MRI brain images and not the gray matter itself.

Table 4 Demographics and clinical characteristics of the study population in the ADNI dataset

| Group | Subject | Age | Gender | |
|---|---|---|---|---|
| | | | M | F |
| AD | 180 | 75.28 ± 7.57 | 98 | 82 |
| sMCI | 105 | 74.69 ± 7.41 | 64 | 41 |
| pMCI | 83 | 73.82 ± 6.65 | 49 | 34 |

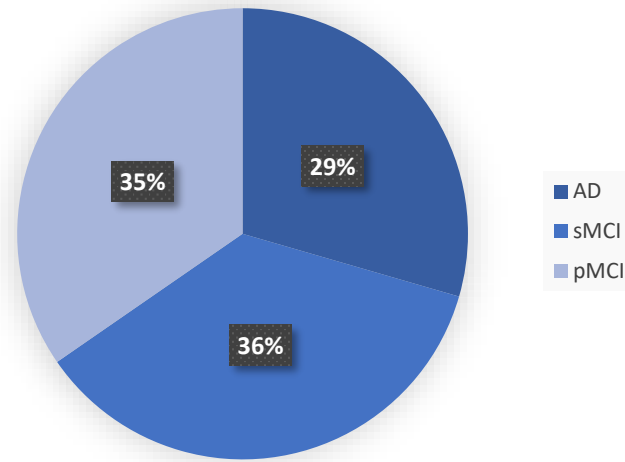

Fig. 9: A chart of the average mini-mental state examination scores of participants across the disease states. The sMCI and pMCI have similar mini-mental scores which are significantly different from AD.

Preprocessing

Considering that structural changes in the brain and AD pathology are more prominent in the sagittal plane, their efficacy in AD classification has long been established[7,12]. This study explored two sagittal plane views, the midsagittal and parasagittal viewpoints, from the ADNI volumetric data in the sagittal plane. From the 3D structured focused ADNI dataset comprising 368 subjects with 155 slices, only slices showing full brain parts were retained; as such, the number of slices varied from sample to sample. In this study, we considered a different approach, which is an unusual practice in the literature where the disease case of a sample is described by all/fraction of their slices. The slices from the sMCI and pMCI patients were combined and referred to as the MCI set, while the slices from the AD samples were likewise combined into an AD set. We consider AD diagnosis to be a sample-independent task; therefore, it is only logical to teach the deep learning algorithm to capture the factors influencing the two categories, MCI and AD, to better understand the disparity that exists between them. In all, a total of 4228 midsagittal AD and MCI sets were retained, each consisting of 2160 and 2068 participants, respectively. The para-sagittal AD and MCI sets included 2700 and 2820 participants, respectively, for a total of 5520 data points.

Despite the intensity correction on the original ADNI dataset, the structure-focused ADNI dataset shows that images are captured under poor illumination conditions. Therefore, we propose to expand the brightness of pixel intensities of the structure-focused ADNI data using the Gamma correction technique. We found the Gamma correction technique to be useful for this task because the nonlinearity property of the Gamma function is better suited for expanding the brightness of pixel intensities of poorly illuminated images without losing information useful for further processing[35]. The gamma correction technique is formed from the inverse of the gamma function. A Gamma value of 0.2 is the value that achieved optimal performance in the proposed framework. The AD and MCI samples were subsequently split into training and testing sets at a ratio of 80:20. The testing set is further split into halves to accommodate the set for validation.

The SNeurodCNN Architecture

As previously stated, each structure-focused ADNI MRI slice is characterized by the formation of gray matter, which forms the anatomical structures of the MR brain images. This structural formation, as opposed to gray matter features, is used to capture focal structural changes associated with neurodegeneration in the cerebral cortex as opposed to focal fine-grain changes within anatomical structures. For this reason, the SNeurodCNN architecture is designed to be a deep convolutional neural network for modelling the structural neurodegeneration of the brain and discriminating between MCI and AD.

The SNeurodCNN architecture as illustrated in Fig. 10. It consists of downsampling convolutional blocks and fully connected layers and utilizes regularization techniques. The usual hierarchical depth of the convolutional layers in the design of a CNN architecture are necessitated to fully capture the diverse low-level and high-level details in an image. Conventionally, the top of this hierarchy captures local features such as edges, corners, texture, and further down the hierarchy are global features like blob formation. Since the input to SNeurodCNN is the structure focused skull-free ADNI data of less dense spatial information compared to the full skull-free grey matter ADNI data, deeper convolutional blocks are likely to lead to model overfitting because the local and global features already take forms without much depth. As such, only two downsampling convolutional blocks are necessary for extracting feature maps from the structural focused ADNI data which are representative of the focal structural atrophy of the brain cerebral cortex. The first downsampling convolutional block contains a 2D convolutional layer of $3\times 3$ and 32 filter dimension and depth, respectively, with Relu activation function and L2 norm regularization with a weight decay rate of 0.01, and $2\times 2$ max-pooling layer. The second convolutional block contains two 2D convolutional layers of $3\times 3$ and 64 filter dimension and depth, respectively, with each layer having a Relu activation function and L2 norm regularization with a weight decay rate of 0.01, and a $2\times 2$ max-pooling layer. Then, the fully connected (FC) layers map the output of the downsampling convolutional blocks to a 1D feature vector of neurons corresponding to the most significant features. At the FC layers, is a dense network (of 500 hidden neurons) incorporating Relu activation function and L2 norm regularization with a weight decay rate of 0.01, a dropout layer with 50% drop rate, and the finally, a softmax classification layer.

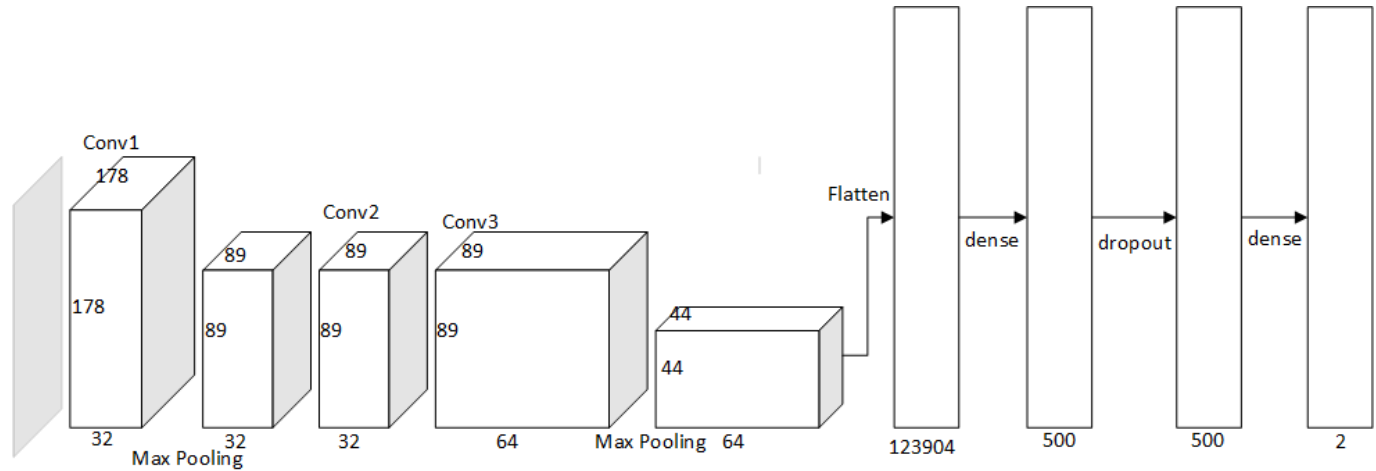

Fig. 10: Architecture of the SNeurodCNN model.

Experimental settings

The network's hyperparameters are chosen in ways that best optimize the network's learning capability. For the hyperparameters, the Adam optimizer with a learning rate fixed at 0.0001 was used. The other variables are epochs and batch sizes, which are 100 and 32, respectively. We adopted early stopping, a regularizer that prevents overfitting by immediately halting training when the performance of the network does not improve after several epochs. The parameters of the early stopping agent used were patience set to 5 and restore_best_weights set to the true Boolean value. During training, the model learns through a method called backpropagation[36,] which updates the weight of the network toward minimizing the gradient error. Considering the impact of hardware resources on model performance, the hardware configuration of the system used for training the network is reported as follows: A Keras library running on a Python-TensorFlow environment with a Google Collab GPU, Tesla K80 (T4).

To evaluate the performance of the SNeurodCNN model, the following metrics were identified and found to be consistent with the metrics used in the literature for diagnosing AD. The variables used were the accuracy, precision, recall, specificity, F1-score, and area under the receiver operating characteristic (ROC) curve (AUC), which provide a comprehensive assessment of the model's performance. Their brief descriptions are as follows.

1. Accuracy: This metric measures the proportion of correct predictions over the total number of predictions, as expressed in (1).

$$Accuracy = \frac{TP + TN}{TP + TN + FP + FN} \qquad (1)$$

where $TP$ = true positive, $TN$ = true negative, $FP$ = false positive and $FN$ = false negative.

2. Precision: the proportion of true positives out of all the instances predicted as positive. In this case, precision measures the number of correctly classified AD cases out of all instances classified as ADs and is mathematically expressed as follows:

$$Precision = \frac{TP}{TP + FP} \qquad (2)$$

3. Recall: also known as sensitivity, this measures the proportion of correctly identified positive instances out of all the actual positive instances. In this case, the recall measures the proportion of correctly classified AD cases out of all actual AD cases. It is given as:

$$Recall = \frac{TP}{TP + FN} \qquad (3)$$

4. Specificity: This metric quantifies the model's ability to make true negative predictions out of all the correctly identified negative instances. With regard to the classification model, specificity will measure the proportion of correctly classified MCI cases out of all actual MCI cases. It is mathematically given as follows:

$$Specificity = \frac{TN}{TN + FP} \qquad (4)$$

5. F1-score: The F1-score combines precision and recall into a single metric, providing a balanced measure of the model's performance. It considers both false positives and false negatives, making it valuable for overall performance assessment.

$$F1 = 2 \times \frac{Precision \times Recall}{Precision + Recall} \quad (5)$$

6. AUROC: The AUROC evaluates the model's performance across different classification thresholds, considering the trade-off between sensitivity and specificity. This approach provides insights into classifier discrimination ability and can be particularly useful for imbalanced datasets.

Limitations

The growing increase in AD cases underscores the pressing necessity of advancing research in deep learning. Our proposed SNeurodCNN achieves remarkable classification accuracies of 98.1% and 97.8% in both the midline and posterior sagittal planes, respectively. This level of performance surpasses that of many existing studies, highlighting the potential of deep learning, particularly our proposed model, in transforming the landscape of neurodegenerative disease diagnosis. The model's sensitivities to neurodegeneration in the brain structure not only substantiated our model's efficacy but also deepened our comprehension of the intricate nature of neurodegenerative diseases in relation to MCI and AD, thereby opening doors for a better understanding of the disease.

While our proposed SNeurodCNN model significantly contributes to AD diagnostic research, its findings are constrained by several limitations. The trainable parameters are large and computationally expensive, especially with respect to reliability and usefulness in real-time clinical diagnosis. Therefore, efforts to achieve high performance while minimizing the computational cost will be explored. Additionally, this paper is directed toward the structural function of the brain analysis, which led to the use of structure-focused ADNIs. However, it will be interesting to understand how the SNeurodCNN model performs on focal fine-grain features available with skull-free ADNIs. This approach enables us to better analyse which of the features, focal fine-grain or focal structure, are representative of the characteristics of neurodegeneration in the brain in a way that makes early diagnosis of AD feasible and reliable in clinical analysis.

We expect that the outcome of the aforementioned study will lead to an understanding of the underlying neurobiological mechanisms involved in MCI and AD, which will help in studying AD progression.

**Conclusion**

In this paper, we propose a structure-focused neurodegeneration convolutional neural network (CNN) architecture called the SNeurodCNN, which was integrated into a deep learning framework along with preprocessing techniques for image enhancement and data preparation. The proposed framework leveraged the midsagittal and parasagittal brain image viewpoints of the structure-focused ADNI dataset. Through experiments, the proposed framework achieved 97.8% accuracy, with 97.0% specificity and 98.5% sensitivity on the parasagittal planes. On the midsagittal plane, accuracy, specificity, and sensitivity of 98.1%, 97.2%, and 99.0%, respectively, were achieved. We further showed that the midsagittal lobe highlights the frontal lobe, occipital lobe, and cerebellum, while the parasagittal lobe extends to the parietal lobe as a region of the brain where structural dynamics are prominent due to MCI and AD. We believe this discovery is useful for identifying digi-biomarkers for the early diagnosis of AD. In future work, efforts will be made to minimize the computational cost of the proposed model while achieving the same level of neurodegeneration modelling. Additionally, it will be interesting to apply the proposed model to focal fine-grained feature learning. This approach enables us to better analyse which of the features, focal fine-grain or focal structure, are representative of the characteristics of neurodegeneration in the brain in a way that makes early diagnosis of AD feasible and reliable in clinical analysis.

Conflict of interest

The authors declare that there are no conflicts of interest.

Code Availability

The source code is available on https://github.com/Simi912/Deep-convolutional-neural-network-classifier-for-Alzheimer-s-Disease-1

Availability of Data

The datasets analysed during the current study are available at the following link: https://doi.gin.g-node.org/10.12751/g-node.aa605a/.

Author contributions

Conceptualization, methodology, model design, and project supervision, C.C.O.; literature review, project scope, K.M, C.C.O, S.O.; data collection, implementation, coding, validation, and testing, S.O., C.C.O; result analysis and interpretation C.C.O, S.O, K.M; manuscript – original draft, C.C.O., K.M, S.O; All authors reviewed, edited, and approved the final manuscript.

To appear in Scientific Report

To appear in Scientific Report18. Zeng, N., Li, H. & Peng, Y. A new deep belief network-based multitask learning for diagnosis of Alzheimer's disease. *Neural Comput Appl* **35**, 11599–11610 (2023).

19. Hazarika, R.A., Abraham, A., Kandar, D. & Maji, A.K. An improved LeNet-deep neural network model for Alzheimer's disease classification using brain magnetic resonance images. *IEEE Access*, *9*, pp.161194-161207, (2021).

20. Hu, Z., Wang, Z., Jin, Y. and Hou, W., VGG-TSwinformer: Transformer-based deep learning model for early Alzheimer's disease prediction. *Computer Methods and Programs in Biomedicine*, *229*, p.107291, (2023).

21. Hazarika, R.A., Maji, A.K., Kandar, D., Jasinska, E., Krejci, P., Leonowicz, Z. and Jasinski, M., 2023. An Approach for Classification of Alzheimer's Disease Using Deep Neural Network and Brain Magnetic Resonance Imaging (MRI). *Electronics*, *12*(3), p.676.

22. Angkoso, C. V., Agustin Tjahyaningtijas, H. P., Purnama, I., & Purnomo, M. H. Multiplane Convolutional Neural Network (Mp-CNN) for Alzheimer's Disease Classification. *International Journal of Intelligent Engineering and Systems* **15**, (2022).

23. Hu, Z., Li, Y., Wang, Z., Zhang, S. & Hou, W. Conv-Swinformer: Integration of CNN and shift window attention for Alzheimer's disease classification. *Comput Biol Med* **164**, 107304 (2023).

24. Bruner, E., Martin-Loeches, M. and Colom, R. Human midsagittal brain shape variation: patterns, allometry and integration. *Journal of Anatomy*, *216*(5), pp.589-599, (2010).

25. Hoffmann, M., 2013. The human frontal lobes and frontal network systems: an evolutionary, clinical, and treatment perspective. *International Scholarly Research Notices*, *2013*.

26. Shi, C., Deng, H., Deng, X., Rao, D. and Yue, W., 2023. The Structural Changes of Frontal Subregions and Their Correlations with Cognitive Impairment in Patients with Alzheimer's Disease. *Journal of Integrative Neuroscience*, *22*(4), p.99.

27. Rehman A, Al Khalili Y. Neuroanatomy, Occipital Lobe. *In: StatPearls. StatPearls Publishing*, PMID: 31335040, (2023).

28. Zhao, L., *et al.* & Alzheimer Disease Neuroimaging Initiative. Risk estimation before progression to mild cognitive impairment and Alzheimer's disease: an AD resemblance atrophy index. *Aging (Albany NY)*, *11*(16), p.6217, (2019).